\documentclass[a4paper]{jpconf}
\usepackage{graphicx}
\usepackage{amssymb}
\usepackage{amsmath}
\def\lsim{\lower -0.3ex \hbox{$<$} \kern -0.75em \lower 0.7ex \hbox{$\sim$}}
\def\gsim{\lower -0.3ex \hbox{$>$} \kern -0.75em \lower 0.7ex \hbox{$\sim$}}

\newcommand{\GVec}[1]{\mbox{\boldmath$#1$}}

\def\Vec#1{{\bf #1}}
\def\GVec#1{\mbox{\boldmath $#1$}}

\def\Journal #1,#2,#3,#4#5#6#7{#1 {\bf #2}, #3 (#4#5#6#7)}

\begin{document}

\title{Interlayer interaction in general incommensurate atomic layers}
\author{Mikito Koshino}
\address{Department of Physics, Tohoku University 
Sendai, 980-8578, Japan
}

\begin{abstract}
We present a general theoretical formulation to describe 
the interlayer interaction in incommensurate bilayer systems 
with arbitrary crystal structures.
By starting from the tight-binding model with the distance-dependent transfer integral,
we show that the interlayer coupling, which is highly complex in the real space,
can be simply written in terms of generalized Umklapp process in the reciprocal space.
The formulation is useful to describe the interaction
in the two-dimensional interface of different materials
with arbitrary lattice structures and relative orientations.
We apply the method to the incommensurate bilayer graphene 
with a large rotation angle, which cannot be treated as a long-range moir\'{e} superlattice,
and obtain the quasi band structure and density of states
within the first-order approximation.
\end{abstract}
\ead{koshino@cmpt.phys.tohoku.ac.jp}

\section{INTRODUCTION}
\label{sec_introduction}

Recently there has been extensive research efforts on atomically-thin nanomaterials, 
including graphene, hexagonal boron nitride(hBN), phosphorene and
the metal transition dichalcogenides.
The hybrid systems composed of different kind of atomic layers
have also attracted much attention,
and there the interaction at the interface between the different atomic layers
often plays an important role in the physical property.
In such a composite system, the lattice periods of the individual atomic layers
are generally incommensurate due to the difference in the crystal structure
and also due to misorientation between the adjacent layers.
A well-known example of irregularly stacked multilayer system
is the twisted bilayer graphene, in which two graphene layers are rotationally stacked
at an arbitrary angle\cite{berger2006electronic}.
When the rotation angle is small, in particular, the system exhibits a moir\'{e} interference 
pattern of which period can be much greater than the atomic scale,
and such a long-period modulation is known to strongly influence the low-energy electronic motion
\cite{lopes2007graphene,mele2010commensuration,trambly2010localization,shallcross2010electronic,morell2010flat,bistritzer2011moirepnas,kindermann2011local,xian2011effects,PhysRevB.86.155449,moon2013opticalabsorption,moon2012energy,moon2013opticalproperties,bistritzer2011moireprb}.
Graphene-hBN composite system
has also been intensively studied as another example of incommensurate 
multilayer system, where the two layers share the same hexagonal lattice structure
but with slightly-different lattice constants, leading to the long-period modulation
even at zero rotation angle
\cite{dean2010boron,xue2011scanning,yankowitz2012emergence,
	ponomarenko2013cloning,hunt2013massive,yu2014hierarchy,yankowitz2014graphene}.
The electronic structure in graphene-hBN system was theoretically studied
\cite{sachs2011adhesion,kindermann2012zero,ortix2012graphene,wallbank2013generic,mucha2013heterostructures,PhysRevB.89.075401,bokdam2014band,jung2013ab,San-Jose2014Spontaneous,song2014topological,uchoa2014valley,neek2014graphene,brey2014coherent,moon2014electronic},
and the recent transport measurements revealed remarkable effects 
such as the formation of mini-Dirac bands
and the fractal subband structure in magnetic fields
\cite{dean2013hofstadter,ponomarenko2013cloning,hunt2013massive,yu2014hierarchy}.

The previous theoretical works mainly targeted the honeycomb lattice
to model twisted bilayer graphene and graphene-hBN system,
and also particularly focus on the long-period moir\'{e} modulation
which arises when the crystal structures of two layers are slightly different.
Then we may ask how to treat general bilayer systems
where the lattice vectors of the adjacent layers are not close to each other.
In this paper, we develop a theoretical formulation to describe
the interlayer interaction effect in general bilayer systems
with arbitrary choice of crystal structures and relative orientations.
By starting from the tight-binding model with distance-dependent transfer integral,
we show that the interlayer coupling
can be simply written in terms of a generalized Umklapp process in the reciprocal space.
We then apply the method to the incommensurate bilayer graphene 
with a large rotation angle ($\theta=20^\circ$) 
which cannot be treated as a long-range moir\'{e} superlattice,
and obtain the quasi band structure and density of states
within the first-order approximation.
Finally, we apply the formulation
to the moir\'{e} superlattice where the two lattice structures are close,
and demonstrate that the long-range effective theory is naturally derived.

\section{Interlayer Hamiltonian for general incommensurate atomic layers}
\label{sec_interlayer}

\begin{figure}
	\begin{center}
		\leavevmode\includegraphics[width=0.5\hsize]{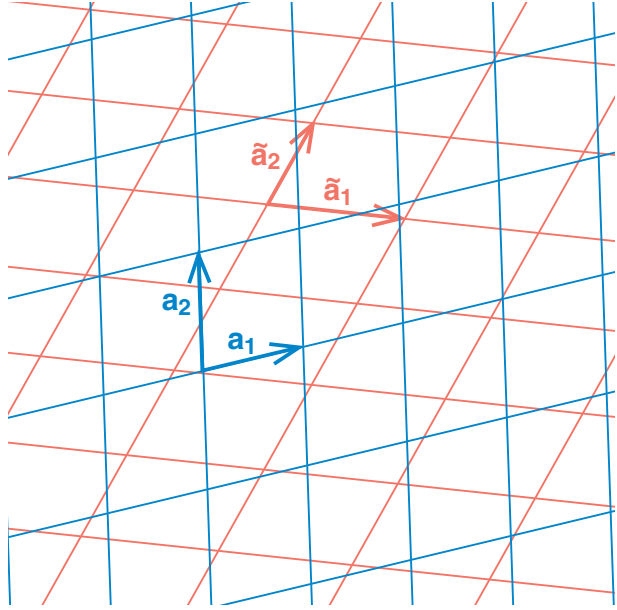}
	\end{center}
	\caption{
		Bilayer system composed of a pair of two-dimensional atomic layers with generally different crystal structures.
		Each unit cell can be generally composed of different sublattices and/or atomic orbitals.
	}
	\label{fig_schem}
\end{figure}

We consider a bilayer system composed of a pair of two-dimensional atomic layers having generally different crystal structures as shown in Fig. \ref{fig_schem}.
We write the primitive lattice vectors as $\Vec{a}_1$ and $\Vec{a}_2$ for layer 1
and $\tilde{\Vec{a}}_1$ and $\tilde{\Vec{a}}_2$ for layer 2,
which are all along in-plane ($x$-$y$) direction. 
The reciprocal lattice vectors are defined by
$\Vec{G}_i$ and $\tilde{\Vec{G}}_i$ for layer 1 and 2, respectively,
so as to satisfy
$\Vec{a}_i\cdot\Vec{G}_j = \tilde{\Vec{a}}_i\cdot\tilde{\Vec{G}}_j = 2\pi\delta_{ij}$.
The area of the unit cell is given by
$S=|\Vec{a}_1\times\Vec{a}_2|$ and $\tilde{S}=|\tilde{\Vec{a}}_1\times\tilde{\Vec{a}}_2|$
for layer 1 and 2, respectively.
Without specifying any details of the model, 
we can easily show that an electronic state with a Bloch wave vector $\Vec{k}$ on layer 1
and one with $\tilde{\Vec{k}}$ on layer 2 are coupled only when
\begin{align}
\Vec{k}+\Vec{G} = \tilde{\Vec{k}}+\tilde{\Vec{G}},
\label{eq_umklapp}
\end{align}
where $\Vec{G}=m_1\Vec{G}_1+m_2\Vec{G}_2$ and
$\tilde{\Vec{G}}=\tilde{m}_1\tilde{\Vec{G}}_1+\tilde{m}_2\tilde{\Vec{G}}_2$
are reciprocal lattice vectors of layer 1 and 2, respectively.
This is regarded as a generalized Umklapp process between arbitrary misoriented crystals,
and it can be easily understood by considering the wave decomposition in the plain wave basis
as follows.
A Bloch state on layer 1  (say $\phi^{(1)}_\Vec{k}$)  is expressed as a summation of
$e^{i(\Vec{k}+\Vec{G})}$ over the reciprocal vectors $\Vec{G}$, and  
one on layer 2 ($\phi^{(2)}_{\tilde{\Vec{k}}} $)
is expressed as a summation of $e^{i(\tilde{\Vec{k}}+\tilde{\Vec{G}})}$
over $\tilde{\Vec{G}}$.
Also, Hamiltonian of the total system consists of Fourier components of
$\Vec{G}$ and $\tilde{\Vec{G}}$.
As a result, the matrix element $\langle \phi^{(2)}_{\tilde{\Vec{k}}} | H | \phi^{(1)}_\Vec{k} \rangle$
can be non-zero only under the condition Eq.\ (\ref{eq_umklapp}).

In the following, we actually calculate the matrix elements
for generalized Umklapp processes using the tight-binding model.
We assume that a unit cell in each layer contains several atomic orbitals, 
which are specified by $X=A,B,\cdots$ for layer 1
and $\tilde{X}=\tilde{A},\tilde{B},\cdots$ for layer 2.
The lattice positions are given by
\begin{align}
&\Vec{R}_{X}=n_1\Vec{a}_{1}+n_2\Vec{a}_{2}+\GVec{\tau}_X
&(\mbox{layer 1}), \nonumber\\
&\Vec{R}_{\tilde{X}}=\tilde{n}_1\tilde{\Vec{a}}_{1}+\tilde{n}_2\tilde{\Vec{a}}_{2}+\GVec{\tau}_{\tilde{X}}
&(\mbox{layer 2}),
\end{align}
where $n_i$ and $\tilde{n}_i$ are integers,
and  $\GVec{\tau}_X$ and $\GVec{\tau}_{\tilde{X}}$ are the sublattice position
inside the unit cell, which can have in-plane and out-of-plane components.
When the atomic layers are completely planar and stacked with interlayer spacing $d$,
for instance,
we have $\GVec{\tau}_X\cdot\Vec{e}_z = 0$ for layer 1
and $\GVec{\tau}_{\tilde{X}}\cdot\Vec{e}_z = d$ for layer 2,
where $\Vec{e}_z$ is the unit vector in $z$-direction. 

Let us define 
$
|\Vec{R}_{X}\rangle  \equiv \phi_{X}(\Vec{r}-\Vec{R}_X)
$
as the atomic state of the sublattice $X$ localized at $\Vec{R}_{X}$.
The atomic orbital $\phi_{X}$ may be different depending on $X$.
We assume the transfer integral 
from the site 
$\Vec{R}_{X}$ to $\Vec{R}_{\tilde{X}}$
is written as
$-T_{\tilde{X}X}(\Vec{R}_{\tilde{X}} - \Vec{R}_{X})$,
i.e., depending on the relative position $\Vec{R}_{\tilde{X}} - \Vec{R}_{X}$
and also on the sort of atomic orbitals of $X$ and $\tilde{X}$.
The interlayer Hamiltonian to couple the layer 1 and 2 is 
then written as
\begin{eqnarray}
	U = -\sum_{X,\tilde{X}}
	T_{\tilde{X}X}(\Vec{R}_{\tilde{X}} - \Vec{R}_{X})
	|\Vec{R}_{\tilde{X}}\rangle\langle\Vec{R}_{X}| + {\rm h.c.}.
	\label{eq_H_int}
\end{eqnarray}
When the superlattice period is huge, 
constructing Hamiltonian in the real space bases
becomes hard because it requires the relative inter-atom position
for every single combination of the atomic sites of layer 1 and layer 2.
On the other hand, the interlayer coupling is described in a simpler manner 
in the reciprocal space.
We define the Bloch basis of each layer as
\begin{align}
	& |\Vec{k},X\rangle = 
	\frac{1}{\sqrt{N}}\sum_{\Vec{R}_{X}} e^{i\Vec{k}\cdot\Vec{R}_{X}}
	|\Vec{R}_{X}\rangle
    &(\mbox{layer 1}), \nonumber\\
	& |\tilde{\Vec{k}},\tilde{X}\rangle = 
	\frac{1}{\sqrt{\tilde{N}}}\sum_{\Vec{R}_{\tilde{X}}} e^{i\tilde{\Vec{k}}\cdot\Vec{R}_{\tilde{X}}}
	|\Vec{R}_{\tilde{X}}\rangle
	&(\mbox{layer 2}),
	\label{eq_bloch_base}
\end{align}
where $\Vec{k}$ and $\tilde{\Vec{k}}$ are the two-dimensional
Bloch wave vectors parallel to the layer,
$N(\tilde{N})$ is the number of unit cell 
of layer 1(2) in the total system area $S_{\rm tot}$. 
Although the layer 1 and layer 2 are generally incommensurate,
we assume the system has a large but finite area 
$S_{\rm tot}=NS=\tilde{N}\tilde{S}$
to normalize the wave function.
Then we can show that
the matrix elements of $U$ between Bloch bases
can be written as,
\begin{align}
U_{\tilde{X}X}(\tilde{\Vec{k}},\Vec{k}) 
&\equiv
\langle\tilde{\Vec{k}},\tilde{X}| U	|\Vec{k},X\rangle
\nonumber\\
&=
-\sum_{\Vec{G},\tilde{\Vec{G}}}
{t}_{\tilde{X}X}(\Vec{k}+\Vec{G})
e^{-i\Vec{G}\cdot\mbox{\boldmath \scriptsize $\tau$}_{X}
	+i\tilde{\Vec{G}}\cdot\mbox{\boldmath \scriptsize $\tau$}_{\tilde{X}}}
\, \delta_{\Vec{k}+\Vec{G},\tilde{\Vec{k}}+\tilde{\Vec{G}}},
\label{eq_hkk}
\end{align}
which is non-zero only when the generalized Umklapp condition
Eq.\ (\ref{eq_umklapp}) is satisfied.
Here ${t}(\Vec{q})$ is the in-plane Fourier transform
of the transfer integral defined by
\begin{eqnarray}
{t}_{\tilde{X}X}(\Vec{q}) = 
\frac{1}{\sqrt{S\tilde{S}}} \int
T_{\tilde{X}X}(\Vec{r}+ z_{\tilde{X}X}\Vec{e}_z) 
e^{-i \Vec{q}\cdot \Vec{r}} d^2r,
\label{eq_ft}
\end{eqnarray}
where $z_{\tilde{X}X} = (\GVec{\tau}_{\tilde{X}}-\GVec{\tau}_{X})\cdot\Vec{e}_z$, 
and the integral in $\Vec{r}$
is taken over the two-dimensional space of $S_{\rm tot}$.
$\Vec{G}$ and $\tilde{\Vec{G}}$ run over all the reciprocal lattice vectors
of layer 1 and 2, respectively.
Since ${t}_{\tilde{X}X}(\Vec{k})$ decays in large $k$,
we only have a limited number of relevant components in the summation of Eq.\ (\ref{eq_hkk}).

Eq.\ (\ref{eq_hkk}) is derived in a straightforward manner as follows.
By inserting $U$ in Eq.\ (\ref{eq_H_int}) to the definition of $U_{\tilde{X}X}$,
we have
\begin{align}
U_{\tilde{X}X}(\tilde{\Vec{k}},\Vec{k}) 
&=
-\frac{1}{\sqrt{N\tilde{N}}}
	\sum_{\Vec{R}_X,\Vec{R}_{\tilde{X}}}
	T_{\tilde{X}X}(\Vec{R}_{\tilde{X}}-\Vec{R}_{X})
	e^{i\Vec{k}\cdot\Vec{R}_X-i\tilde{\Vec{k}}\cdot\Vec{R}_{\tilde{X}}}
	\nonumber\\	
&=
	-\frac{1}{\sqrt{N\tilde{N}}}
	\sum_{\Vec{R}_X}
	e^{i(\Vec{k}-\tilde{\Vec{k}})\cdot\Vec{R}_X}
	\sum_{\Vec{R}_{\tilde{X}}}
 T_{\tilde{X}X}(\Vec{R}_{\tilde{X}}-\Vec{R}_{X})
	e^{-i\tilde{\Vec{k}}\cdot(\Vec{R}_{\tilde{X}}-\Vec{R}_X)}
	\label{eq_hkk2}
\end{align}
By applying the inverse Fourier transform 
of Eq.\ (\ref{eq_ft}),
\begin{eqnarray}
T_{\tilde{X}X}(\Vec{r}+ z_{\tilde{X}X}\Vec{e}_z)  = 
				\frac{1}{\sqrt{N\tilde{N}}} \int
				{t}_{\tilde{X}X}(\Vec{q}) 
				e^{i \Vec{q}\cdot \Vec{r}} d^2q,
\label{eq_inv_ft}
\end{eqnarray}
to Eq.\ (\ref{eq_hkk2}), the second summation is transformed as
\begin{align}
\sum_{\Vec{R}_{\tilde{X}}}
&T_{\tilde{X}X}(\Vec{R}_{\tilde{X}}-\Vec{R}_{X})
e^{-i\tilde{\Vec{k}}\cdot(\Vec{R}_{\tilde{X}}-\Vec{R}_X)}
\nonumber\\
&=
\frac{1}{\sqrt{N\tilde{N}}} \int d^2q \,
e^{i(\Vec{q}-\tilde{\Vec{k}})\cdot
	(\mbox{\boldmath \scriptsize $\tau$}_{\tilde{X}}-\Vec{R}_X)}
\,
{t}_{\tilde{X}X}(\Vec{q}) 
\sum_{\tilde{n}_1 \tilde{n}_2}
e^{i (\Vec{q}-\tilde{\Vec{k}})\cdot (\tilde{n}_1\tilde{\Vec{a}}_{1}+\tilde{n}_2\tilde{\Vec{a}}_{2})}
\nonumber\\
&=
\sqrt{\frac{\tilde{N}}{N}} 
\sum_{\tilde{\Vec{G}}}
{t}_{\tilde{X}X}(\tilde{\Vec{k}}+\tilde{\Vec{G}}) 
e^{i \tilde{\Vec{G}}\cdot 
	(\mbox{\boldmath \scriptsize $\tau$}_{\tilde{X}}-\Vec{R}_X)},
\label{eq_hkk3}
\end{align}
where we used in the last equation, 
\begin{align}
\sum_{\tilde{n}_1 \tilde{n}_2}
e^{i (\Vec{q}-\tilde{\Vec{k}})\cdot (\tilde{n}_1\tilde{\Vec{a}}_{1}+\tilde{n}_2\tilde{\Vec{a}}_{2})}
= \tilde{N} \sum_{\tilde{\Vec{G}}}\delta_{\Vec{q}-\tilde{\Vec{k}},\tilde{\Vec{G}}}.
\end{align}
Using Eqs.\ (\ref{eq_hkk2}) and (\ref{eq_hkk3}),
we have
\begin{align}
U_{\tilde{X}X}(\tilde{\Vec{k}},\Vec{k}) 
&=
-\frac{1}{N}
\sum_{\tilde{\Vec{G}}}{t}_{\tilde{X}X}(\tilde{\Vec{k}}+\tilde{\Vec{G}}) 
e^{i \tilde{\Vec{G}}\cdot 
	\mbox{\boldmath \scriptsize $\tau$}_{\tilde{X}}}
\sum_{\Vec{R}_X}
e^{i (\Vec{k}-\tilde{\Vec{k}}-\tilde{\Vec{G}})\cdot\Vec{R}_{X}}
\nonumber\\
&=
-\sum_{\Vec{G},\tilde{\Vec{G}}}
{t}_{\tilde{X}X}(\Vec{k}+\Vec{G})
e^{-i\Vec{G}\cdot\mbox{\boldmath \scriptsize $\tau$}_{X}
	+i\tilde{\Vec{G}}\cdot\mbox{\boldmath \scriptsize $\tau$}_{\tilde{X}}}
\, \delta_{\Vec{k}+\Vec{G},\tilde{\Vec{k}}+\tilde{\Vec{G}}},
\label{eq_hkk4}
\end{align}
which is Eq.\ (\ref{eq_hkk}). 
In the summation in $\Vec{R}_X$ in the first line,
we used the transformation 
\begin{align}
\sum_{\Vec{R}_X}
&e^{i (\Vec{k}-\tilde{\Vec{k}}-\tilde{\Vec{G}})\cdot\Vec{R}_{X}}
=
\sum_{n_1 n_2}
e^{i (\Vec{k}-\tilde{\Vec{k}}-\tilde{\Vec{G}})\cdot
	(n_1\Vec{a}_{1}+n_2\Vec{a}_{2}+
	\mbox{\boldmath \scriptsize $\tau$}_{X})}
\nonumber\\
&=
e^{i (\Vec{k}-\tilde{\Vec{k}}-\tilde{\Vec{G}})\cdot
		\mbox{\boldmath \scriptsize $\tau$}_{X}}
\sum_{n_1 n_2}
e^{i (\Vec{k}-\tilde{\Vec{k}}-\tilde{\Vec{G}})\cdot
	(n_1\Vec{a}_{1}+n_2\Vec{a}_{2})}
\nonumber\\
&=
e^{i (\Vec{k}-\tilde{\Vec{k}}-\tilde{\Vec{G}})\cdot
	\mbox{\boldmath \scriptsize $\tau$}_{X}}
N\sum_{\Vec{G}}
\delta_{\Vec{k}-\tilde{\Vec{k}}-\tilde{\Vec{G}},-\Vec{G}}.
\label{eq_hkk5}
\end{align}

\section{Irregularly stacked honeycomb lattices}
\label{sec_irr}

\begin{figure}
	\begin{center}
		\leavevmode\includegraphics[width=1.\hsize]{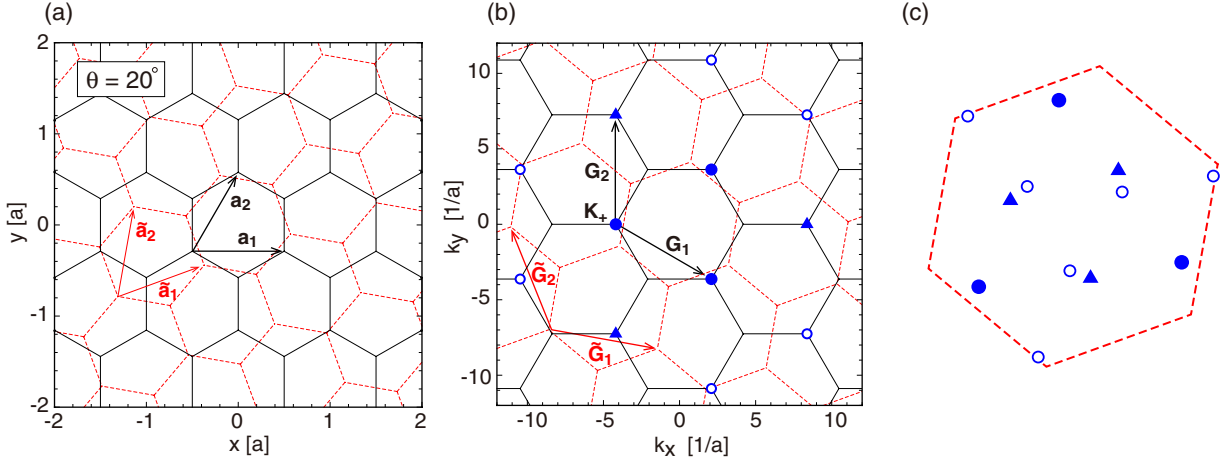}
	\end{center}
	\caption{
		(a)  Irregularly stacked bilayer graphene
		at rotation angle $\theta=20^\circ$.
		(b) Brillouin zones
		of the individual layers in the extended zone scheme.
		Symbols (filled circles, triangles and open circles)
		represent the positions of $\Vec{K}_+ +\Vec{G}$
		with several $\Vec{G}$'s.
		The different symbols indicate the different distances from the origin.
		(c) Corresponding positions of the symbols in (b)
		inside the first Brillouin zone of layer 2.
	}
	\label{fig_honeycomb_20}
\end{figure}

We apply the general formulation obtained above
to the irregularly stacked bilayer graphene with an arbitrary rotation angle.
Here we consider a pair of hexagonal lattices as shown in Fig.\ \ref{fig_honeycomb_20}(a),
which are stacked with a relative rotation angle $\theta$
and interlayer spacing $d$.
We define the primitive lattice vectors 
of layer 1 as  $\Vec{a}_1 = a(1,0)$ and $\Vec{a}_2 = a(1/2,\sqrt{3}/2)$ 
with the lattice constant $a$,
and those of the layer 2 as
$\tilde{\Vec{a}}_i = R\,\Vec{a}_i$ where
$R$ is the rotation matrix by $\theta$.
Accordingly, the reciprocal lattice vectors
of layer 1 ($\Vec{G}_1, \Vec{G}_2$)
and layer 2 are related by
$\tilde{\Vec{G}}_i = R\, \Vec{G}_i$.
The atomic positions are given by
\begin{align}
&\Vec{R}_{X}=n_1\Vec{a}_{1}+n_2\Vec{a}_{2}+\GVec{\tau}_X
&(\mbox{layer 1}), \nonumber\\
&\Vec{R}_{\tilde{X}}=\tilde{n}_1\tilde{\Vec{a}}_{1}+\tilde{n}_2\tilde{\Vec{a}}_{2}+\GVec{\tau}_{\tilde{X}}
&(\mbox{layer 2}),
\end{align}
for $X=A,B$ (layer 1) and  $\tilde{X}=\tilde{A},\tilde{B}$ (layer 2), where
\begin{align}
&\GVec{\tau}_{A}= 0, \nonumber\\
&\GVec{\tau}_{B}= -(\Vec{a}_1+2\Vec{a}_2)/3,  \nonumber\\
&\GVec{\tau}_{\tilde{A}}= d\,\Vec{e}_z+ \GVec{\tau}_0, \nonumber\\
&\GVec{\tau}_{\tilde{B}}= d\,\Vec{e}_z+\GVec{\tau}_0-(\tilde{\Vec{a}}_1+2\tilde{\Vec{a}}_2)/3.
\end{align}
Here we take the origin at an $A$ site,
and define $\GVec{\tau}_0$ as the relative in-plane translation vector of the layer 2 to layer 1.


To describe the electron's motion,
we adopt the single-orbital tight-binding model for $p_z$ atomic orbitals.
Then $T_{\tilde{X}X}(\Vec{R})$ does not depend on indexes $\tilde{X}$ and $X$,
and it is approximately written 
in terms of the Slater-Koster parametrization as, \cite{slater1954simplified}
\begin{eqnarray}
&& -T(\Vec{R}) = 
V_{pp\pi}\left[1-\left(\frac{\Vec{R}\cdot\Vec{e}_z}{d}\right)^2\right]
+ V_{pp\sigma}\left(\frac{\Vec{R}\cdot\Vec{e}_z}{d}\right)^2,
\nonumber \\
&& V_{pp\pi} =  V_{pp\pi}^0 e^{- (R-a/\sqrt{3})/r_0},
\quad V_{pp\sigma} =  V_{pp\sigma}^0  e^{- (R-d)/r_0}.
\nonumber\\ 
\label{eq_transfer_integral}
\end{eqnarray}
The typical parameters for graphene are
$a\approx 0.246$nm, $d\approx 0.335\,\mathrm{nm}$,
$V_{pp\pi}^0 \approx -2.7\,\mathrm{eV}$,
$ V_{pp\sigma}^0 \approx 0.48\,\mathrm{eV}$,
and $r_0 \approx 0.184 a$\cite{moon2013opticalabsorption}.
Once the transfer integral $T(\Vec{R})$ between the atomic sites
is given, one can calculate the in-plane Fourier transform $t(\Vec{q})$, 
and then obtain the interlayer Hamiltonian by Eq.\ (\ref{eq_hkk}).
Since the transfer integral is isotropic along the in-plane direction,
we can write $t(\Vec{q}) = t(q)$ with $q=|\Vec{q}|$.

Fig.\ \ref{fig_honeycomb_20}(a) illustrates
the lattice structure at rotation angle $\theta=20^\circ$.
Fig.\ \ref{fig_honeycomb_20}(b) shows the Brillouin zones
of the two layers in the extended zone scheme,
where the blue symbols (filled circles, triangles and open circles)
represent the positions of $\Vec{k} +\Vec{G}$
for some particular $\Vec{k}$ (here chosen as the zone corner $\Vec{K}_+$)
with several $\Vec{G}$'s.
Figs.\ \ref{fig_honeycomb_20}(c) plots 
the corresponding positions of these symbols inside the first Brillouin zone of layer 2.
This indicates the wave numbers of layer 2, 
$\tilde{\Vec{k}} = \Vec{k} +\Vec{G} - \tilde{\Vec{G}}$,
which are coupled to $\Vec{k}$ of layer 1 under the condition of Eq.\ (\ref{eq_umklapp}).
The amplitude of the coupling is given by $t(\Vec{k} +\Vec{G})$,
and it solely depends on 
the distance to each symbol from the $k$-space origin in Fig.\ \ref{fig_honeycomb_20}(b).
In the present parameter choice, 
the amplitudes for filled circles, triangles and open circles
are $t(K)\approx 110$ meV, $t(2K)\approx 1.6$ meV,
and $t(\sqrt{7}K)\approx 0.062$ meV respectively,
where $K=|\Vec{K}_+| = 4\pi/(3a)$.
The couplings for other $k$-points are exponentially small and negligible.

When the lattice vectors of the two layers are incommensurate
(i.e., do not have a common multiple) as in this example,
we cannot define the common Brillouin zone nor calculate 
the exact band structure,
since the interlayer interaction connects infinite number of $k$-points in
the Brillouin zones of layer 1 and layer 2.
But still, we can obtain an approximate band structure, considering 
only the first-order interlayer processes while neglecting multiple processes. 
Let us consider a particular wave vector $\Vec{k}$ of layer 1,
and take all $\tilde{\Vec{k}}$'s in layer 2 
which are directly coupled to $\Vec{k}$ as in Fig.\ \ref{fig_honeycomb_20}(c). 
By neglecting exponentially small matrix elements,
we can construct a finite Hamiltonian matrix including only the bases
of a single wave vector $\Vec{k}$ of layer 1 and several $\tilde{\Vec{k}}$'s in layer 2.
By diagonalizing the matrix,
we obtain the energy eigenvalues $\varepsilon_{n\Vec{k}}$ 
labeled by the index $n$.
We then define the spectral function contributed from layer 1 as
\begin{equation}
A_1(\Vec{k},\varepsilon)	
= \sum_{n} 
g^{(1)}_{n\Vec{k}} \delta(\varepsilon-\varepsilon_{n\Vec{k}}),
\label{eq_A1}
\end{equation}
where $g^{(1)}_{n\Vec{k}}$ is the total wave amplitudes 
on layer 1 in the state $|n\Vec{k}\rangle$.
The density of states contributed from layer 1 is expressed as
\begin{equation}
D_1(\varepsilon)
=\sum_{\Vec{k}}A_1(\Vec{k},\varepsilon),
\end{equation}
where the summation is taken over the first Brillouin zone of layer 1.
We perform the exactly same procedure for the layer 2
by considering a single $\tilde{\Vec{k}}$ of layer 2
and all $\Vec{k}$'s in layer 1 coupled to $\tilde{\Vec{k}}$,
and obtain the spectral function and the density of states for the layer 2.
The total density of states of the system is given by $D_1+D_2$.

\begin{figure}
	\begin{center}
		\leavevmode\includegraphics[width=0.6\hsize]{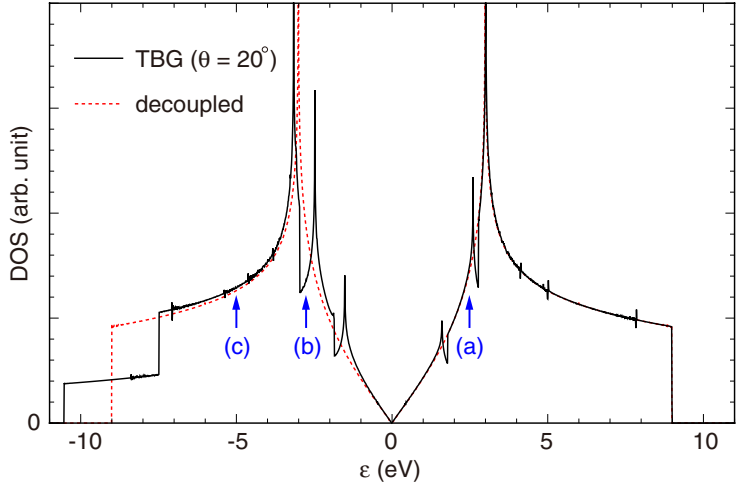}
	\end{center}
	\caption{
		Electronic density of states in the incommensurate bilayer graphene with $\theta=20^\circ$,
		calculated in the first-order approximation (see the text).
		The red dashed curve plots the density of states of decoupled bilayer graphene.
		Vertical arrows indicate the energies considered in Fig.\ \ref{fig_quasi_band}.
	}
	\label{fig_dos}
\end{figure}

\begin{figure}
	\begin{center}
		\leavevmode\includegraphics[width=0.75\hsize]{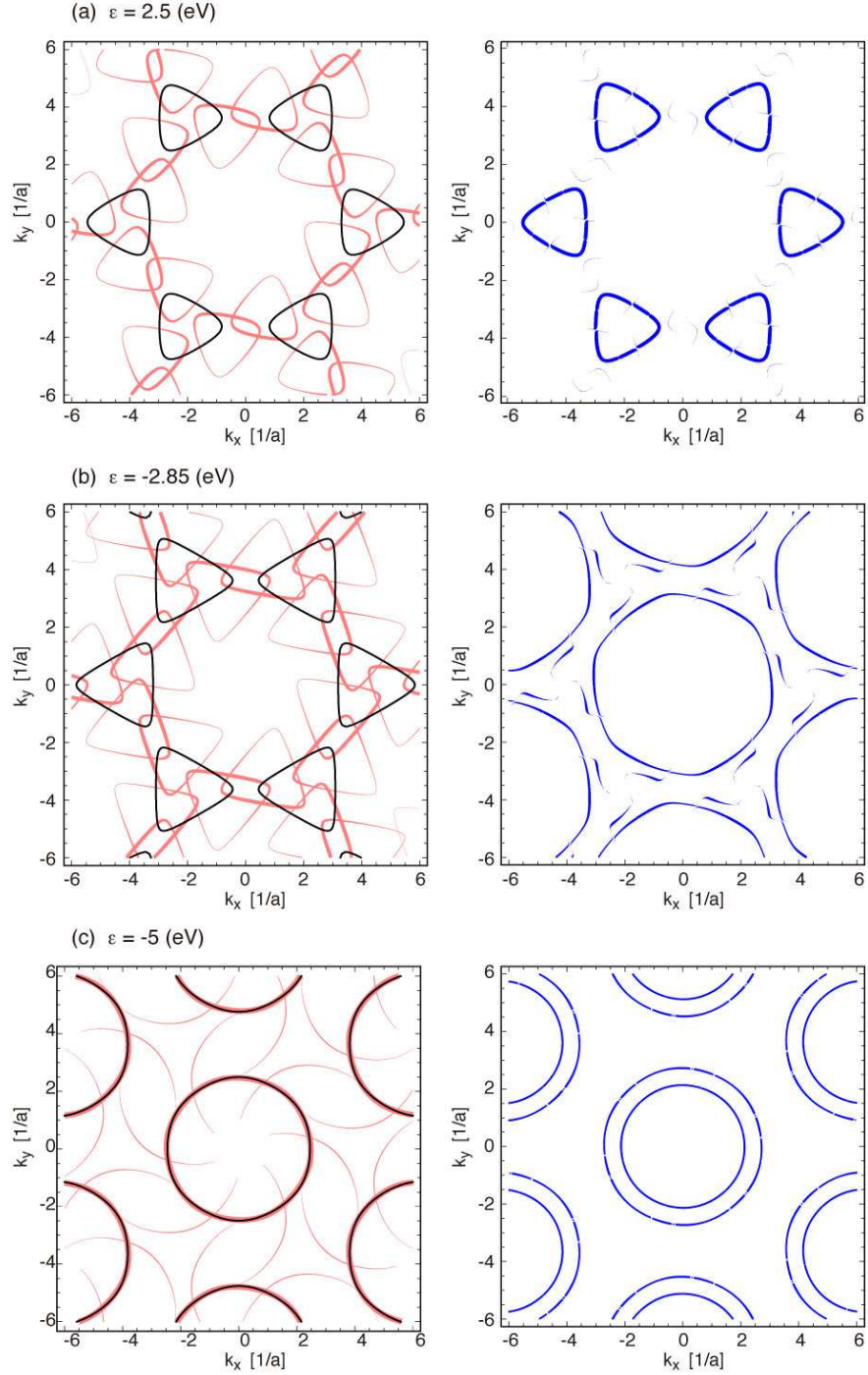}
	\end{center}
	\caption{
		Fermi surface reconstruction in the incommensurate bilayer graphene with $\theta=20^\circ$, at energies of
		(a) $2.5$eV, (b) $-2.85$eV and (c) $-5$eV.
		The right panel in each row shows the layer 1's spectral function $A_1(\Vec{k},\varepsilon)$
		in presence of the interlayer coupling.
		The left panel shows the Fermi surface in absence of the interlayer coupling,
		where the black curve is for the layer 1's energy dispersion $\varepsilon_1(\Vec{k})$,
		and the pink curve is for the layer 2's dispersion with $k$-space shift, $\varepsilon_2(\Vec{k}+\Vec{G})$.
		Thickness of the pink curve indicates the absolute value of the interlayer hopping
		$t(\Vec{k}+\Vec{G}) $.
	}
	\label{fig_quasi_band}
\end{figure}

Fig.\ \ref{fig_dos} plots the total density of state $D(\varepsilon)$
calculated for the incommensurate bilayer graphene with $\theta=20^\circ$.
The red dashed curve is the density of states of decoupled bilayer graphene
(i.e., twice of monolayer's).
We actually see additional peak structures due to the interlayer interaction,
and these features are consistent with the exact tight-binding 
calculation for a commensurate bilayer graphene 
at a similar rotation angle \cite{moon2013opticalabsorption}.
Fig.\ \ref{fig_quasi_band} illustrates the 
Fermi surface reconstruction at several different energies.
The right panel in each row shows the layer 1's spectral function $A_1(\Vec{k},\varepsilon)$
in presence of the interlayer coupling.
The left panel shows the Fermi surface in absence of the interlayer coupling,
where the black curves represent the equienergy lines
for the layer 1's energy dispersion $\varepsilon_1(\Vec{k})$,
and the pink curves are for the layer 2's dispersion with $k$-space shift, $\varepsilon_2(\Vec{k}+\Vec{G})$.
Thickness of the pink curves indicate the absolute value of the interlayer coupling
$t(\Vec{k}+\Vec{G}) $.

Fig.\ \ref{fig_quasi_band}(a) shows a typical case ($\varepsilon=2.5$ eV)
where we observe small band anticrossing 
at the intersection of the Fermi surfaces of layer 1 and layer 2.
Fig.\ \ref{fig_quasi_band}(b) ($\varepsilon=-2.85$ eV) is for the energy at which 
the density of states exhibits a dip [Fig.\ \ref{fig_dos}] .
There the interlayer coupling is relatively strong,
and indeed we see that the original triangular Fermi pockets of the individual layers
are strongly mixed and reconstructed into a large single Fermi surface.
Note that the coupling strength depends not only on $t(\Vec{k}+\Vec{G}) $, but also
the relative phase factors between different sublattices.
At even lower energy $\varepsilon=-5$ eV [Fig.\ \ref{fig_quasi_band}(b)]
beyond the van-Hove singularity,
the layer 1 and the layer 2 have almost identical Fermi surface
surrounding $\Gamma$ point in absence of the interlayer coupling,
and they are hybridized into a pair of circles with different radii
corresponding to the bonding and anti-bonding states.
This feature is also observed in the density of states [Fig.\ \ref{fig_dos}]
as a large split of the band bottom.
Such a splitting does not occur in the positive energy region,
because there $A$ and $B$ sublattices in the same layer
have the opposite signs in the wave amplitude,
so that the interlayer mixing vanishes due to the phase cancellation.

In the above approximation, we neglect the second order processes
such that $\tilde{\Vec{k}}$ points of layer 2 (linked from initial $\Vec{k}$ of layer 1) 
are further coupled to other $\Vec{k}'$ of layer 1.
We can include such higher-order processes up to any desired order as follows.
To include the second order process in calculating $A_1(\Vec{k},\varepsilon)$, for example,
we take all $\tilde{\Vec{k}}$ of layer 2 which are coupled to $\Vec{k}$ of layer 1
and also take all $\Vec{k}'$ of layer 1 which are coupled to $\tilde{\Vec{k}}$,
and then construct a Hamiltonian matrix with a larger dimension.
From the obtained eigenstates, we calculate the spectral function
using Eq.\ (\ref{eq_A1}), but then $g^{(1)}_{n\Vec{k}}$ should be the total wave amplitudes 
from layer 1 at the original $\Vec{k}$ (0-th order), without including $\Vec{k}'$.

\begin{figure}
	\begin{center}
		\leavevmode\includegraphics[width=1.\hsize]{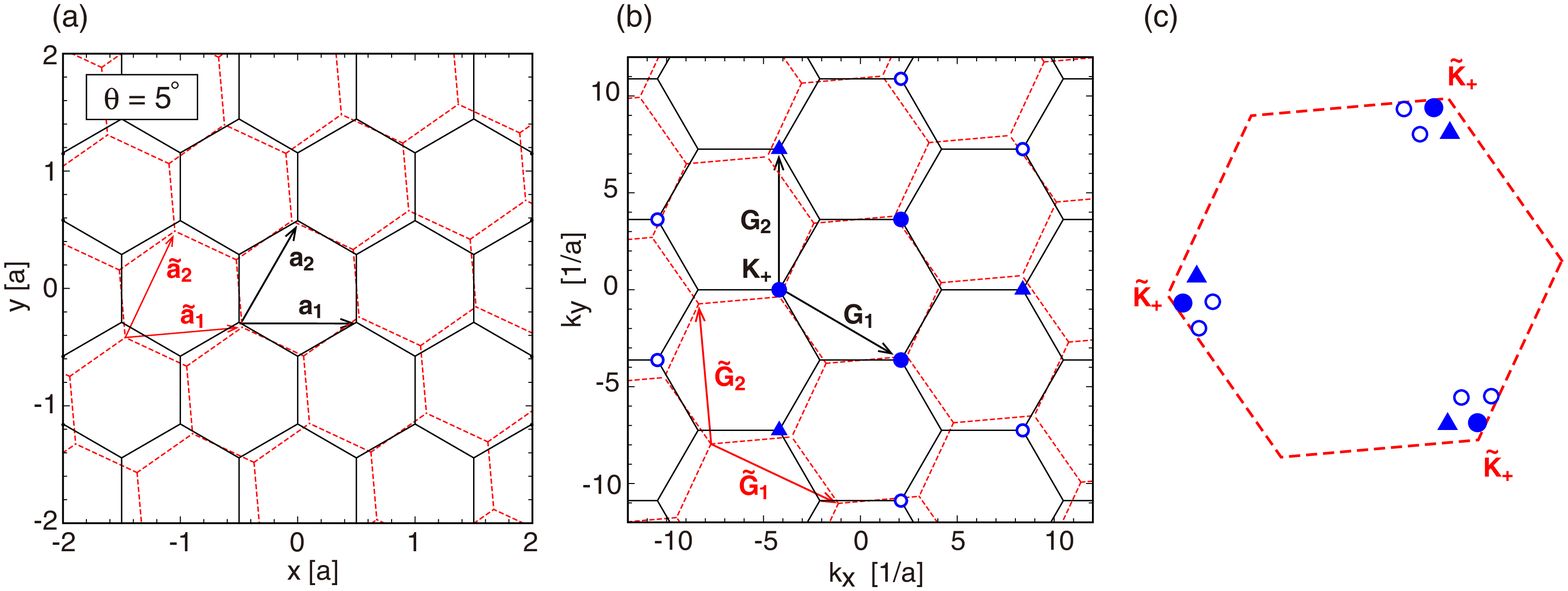}
	\end{center}
	\caption{
		Similar plots to Fig. \ref{fig_honeycomb_20} 
		for $\theta=5^\circ$.
	}
	\label{fig_honeycomb_5}
\end{figure}

\section{Long-period moir\'{e} superlattice}
\label{sec_moire}

When the primitive lattice vectors of layer 1 
and those of layer 2 are slightly different,
the interference of two lattice structures
give rise to a long-period moir\'{e} pattern,
and then we can use the long-range effective theory
to describe the interlayer interaction.
\cite{lopes2007graphene,mele2010commensuration,trambly2010localization,shallcross2010electronic,morell2010flat,bistritzer2011moirepnas,kindermann2011local,xian2011effects,PhysRevB.86.155449,moon2013opticalabsorption,moon2012energy,moon2013opticalproperties,bistritzer2011moireprb}
Here we show that the long-range effective theory
can be naturally derived from the present 
general formulation, just by assuming that the two lattice structures are close to each other.

We define a linear transformation with a matrix $A$
that relates the primitive lattice vectors of layer 1 and 2 as
\begin{align}
\tilde{\Vec{a}}_i = A\,\Vec{a}_i.
\end{align} 
Correspondingly, the reciprocal lattice vectors become
\begin{align}
\tilde{\Vec{G}}_i = (A^\dagger)^{-1}\Vec{G}_i,
\end{align} 
to satisfy $\Vec{a}_i\cdot\Vec{G}_j = \tilde{\Vec{a}}_i\cdot\tilde{\Vec{G}}_j = 2\pi \delta_{ij}$.
When the lattice structures of the two layers are similar,
the matrix $A$ is close to the unit matrix.
The reciprocal lattice vectors of the moir\'{e}
superlattice is given by small difference between $\Vec{G}_i$ and $\tilde{\Vec{G}}_i$ as
\begin{align}
	\Vec{G}^M_i &= \Vec{G}_i - \tilde{\Vec{G}}_i = [1-(A^\dagger)^{-1}]\Vec{G}_i.
	\label{eq_GM}
\end{align} 
When the two layers are identical and rotationally stacked with a small angle, for example,
the matrix $A$ is given by a rotation $R$.
Noting $(R^\dagger)^{-1}=R$, we have
\begin{align}
&\Vec{G}^M_i = (1-R)\Vec{G}_i.
\end{align} 
When layer 1 and layer 2 have different lattice constant as in the graphene-hBN bilayer,
the matrix $A$ is given by the combination of the isotropic expansion 
and the rotation as $\alpha R$, where $\alpha$ is the lattice constant ratio.
Since $[(\alpha R) ^\dagger]^{-1}=\alpha^{-1} R$, we have
\begin{align}
&\Vec{G}^M_i = [1-\alpha^{-1} R]\Vec{G}_i.
\end{align} 
When the specific form of the matrix $A$ is given,
we can immediately derive the interlayer matrix elements 
for the long wavelength components
using Eq.\ (\ref{eq_hkk}) as,
\begin{align}
	U_{\tilde{X}X}(\Vec{k}+m_1\Vec{G}^M_1+m_2\Vec{G}^M_2,\Vec{k}) 
	&=
	-{t}_{\tilde{X}X}(\Vec{k}+m_1\Vec{G}_1+m_2\Vec{G}_2)
	\nonumber\\
	&\qquad\times 
	\, e^{-i(m_1\Vec{G}_1+m_2\Vec{G}_2)
		\cdot\mbox{\boldmath \scriptsize $\tau$}_{X}
		+i(m_1\tilde{\Vec{G}}_1+m_2\tilde{\Vec{G}}_2)
		\cdot\mbox{\boldmath \scriptsize $\tau$}_{\tilde{X}}},
	\label{eq_hkk_moire}
\end{align}
where $m_1$ and $m_2$ are integers.

For example, let us derive the interlayer Hamiltonian of the irregularly stacked 
bilayer graphene with a small rotation angle. 
Since the low-energy spectrum of graphene is dominated by the electronic states around
the Brillouin zone corners $K$ and $K'$, 
we consider the matrix elements for initial and final $k$-vectors near those points.
The $K$ and $K'$ points are located at $\Vec{K}_\xi = -\xi (2\Vec{G}_1+\Vec{G}_2)/3$ for layer 1
and $\tilde{\Vec{K}}_\xi = -\xi (2\tilde{\Vec{G}}_1+\tilde{\Vec{G}}_2)/3$ for layer 2,
where $\xi=\pm 1$ are the valley indexes for $K$ and $K'$, respectively.
When we start from $\Vec{k}=\Vec{K}_+$, for example,
a typical scattering process is illustrated
in Figs.\ \ref{fig_honeycomb_5}(b) and (c)
in a similar manner to Fig.\ \ref{fig_honeycomb_20}.
There the electron at $\Vec{K}_+$ in layer 1
is coupled to $\Vec{K}_+ +m_1\Vec{G}_1+m_2\Vec{G}_2$
in layer 2 with absolute amplitude $t(\Vec{K}_+ +m_1\Vec{G}_1 + m_2+\Vec{G}_2)$.
As already argued in the previous section,
the coupling amplitudes for filled circles, triangles and open circles
in Figs.\ \ref{fig_honeycomb_5}(b) and (c)
are $t(K)\approx 110$ meV, $t(2K)\approx 1.6$ meV,
and $t(\sqrt{7}K)\approx 0.062$ meV respectively,
and the couplings to other $k$-points are negligibly small.
The matrix element changes when the initial vector $\Vec{k}$ is shifted from $\Vec{K}_+$,
but we neglect such a dependence assuming $\Vec{k}$ is close to $\Vec{K}_+$.
As a result, we obtain
the interlayer Hamiltonian of near $\Vec{K}_\xi$ from Eq.\ (\ref{eq_hkk_moire}) as
\begin{align}
 U &= 
\begin{pmatrix}
U_{\tilde{A} A} & U_{\tilde{A} B}
\\
U_{\tilde{B} A} & U_{\tilde{B} B}
\end{pmatrix}
\nonumber\\
&
=
t(K)
\Biggl[
\begin{pmatrix}
1 & 1
\\
1 & 1
\end{pmatrix}
+
\begin{pmatrix}
1 & \omega^{-\xi}
\\
\omega^{\xi} & 1
\end{pmatrix}
e^{i\xi\Vec{G}^{\rm M}_1\cdot\Vec{r}}
+
\begin{pmatrix}
1 & \omega^{\xi}
\\
\omega^{-\xi} & 1
\end{pmatrix}
e^{i\xi(\Vec{G}^{\rm M}_1+\Vec{G}^{\rm M}_2)\cdot\Vec{r}}
\Biggr]
\nonumber\\
&+
t(2K)
\Biggl[
\begin{pmatrix}
1 & 1
\\
1 & 1
\end{pmatrix}
e^{i\xi(2\Vec{G}^{\rm M}_1+\Vec{G}^{\rm M}_2)\cdot\Vec{r}}
+
\begin{pmatrix}
1 & \omega^{-\xi}
\\
\omega^{\xi} & 1
\end{pmatrix}
e^{i\xi\Vec{G}^{\rm M}_2\cdot\Vec{r}}
+
\begin{pmatrix}
1 & \omega^{\xi}
\\
\omega^{-\xi} & 1
\end{pmatrix}
e^{-i\xi\Vec{G}^{\rm M}_2\cdot\Vec{r}}
\Biggr]
\nonumber\\
&+\cdots,
\label{eq_U}
\end{align}
where $\Vec{r}$ is the in-plane position, 
$\omega = \exp(2\pi i/3)$ and $\GVec{\tau}_0$ is set to $0$.

The total Hamiltonian is written in the basis of $\{A,B,\tilde{A},\tilde{B} \}$ as
\begin{eqnarray}
	U^{(\xi)}_{\rm eff} = 
	\begin{pmatrix}
		H_1 & U^\dagger \\
		U & H_2
	\end{pmatrix},
	\label{eq_eff_model}
\end{eqnarray}
where
$H_1$ and $H_2$ are the intralayer Hamiltonian of layer 1 and 2, respectively, 
defined by
\begin{eqnarray}
	&& H_1 \approx -\hbar v ({\Vec{k}}-\Vec{K}_\xi)
	\cdot (\xi \sigma_x, \sigma_y), \nonumber\\
	&& H_2 \approx -\hbar v [R^{-1}({\Vec{k}}-\tilde{\Vec{K}}_\xi)]
	\cdot (\xi \sigma_x, \sigma_y),
\end{eqnarray}
with  Pauli matrices $\sigma_x$ and $\sigma_y$,
and the graphene's band velocity $v$.
If we only take the terms with $t(K)$ in the matrix $U$, 
the expression becomes consistent with the previous formulation
for graphene-graphene bilayer. 
\cite{lopes2007graphene,bistritzer2011moirepnas,kindermann2011local,moon2013opticalabsorption}
Note that the Hamiltonian matrix
depends on the actual choice of $K$ points out of the equivalent 
Brillouin zone corners; $\Vec{K}_{\xi} + m_1 \Vec{G}_1 + m_2\Vec{G}_2$ 
and $\tilde{\Vec{K}}_{\xi} + \tilde{m}_1 \tilde{\Vec{G}}_1 + \tilde{m}_2 \tilde{\Vec{G}}_2$.
The different choice of $m_i$ and $\tilde{m}_i$ adds extra phase factors to
the Bloch bases depending on sublattices, while the resulting Hamiltonian matrix
are connected to the original by an unitary transformation.

While we neglected the interlayer translation vector $\GVec{\tau}_0$ 
in deriving Eq.\ (\ref{eq_U}), the matrix elements of $U$ actually depend
on $\GVec{\tau}_0$ according to Eq.\ (\ref{eq_hkk_moire}), 
where the term with $e^{i(m_1\Vec{G}^{\rm M}_1+m_2\Vec{G}^{\rm M}_2)\cdot\Vec{r}}$
is accompanied by an additional phase factor
$e^{i(m_1\tilde{\Vec{G}}_1+m_2\tilde{\Vec{G}}_2)
	\cdot\mbox{\boldmath \scriptsize $\tau$}_{0}}$.
This extra term, however, can be incorporated into a shift of the space origin as
\begin{align}
e^{i(m_1\Vec{G}^{\rm M}_1+m_2\Vec{G}^{\rm M}_2)\cdot\Vec{r}}\, e^{i(m_1\tilde{\Vec{G}}_1+m_2\tilde{\Vec{G}}_2)
	\cdot\mbox{\boldmath \scriptsize $\tau$}_{0}}
= e^{i(m_1\Vec{G}^{\rm M}_1+m_2\Vec{G}^{\rm M}_2)\cdot(\Vec{r}+\Vec{r}_0)},
\end{align}
where $\Vec{r}_0 = (A-1)\GVec{\tau}_0$.
This reflects the fact that relative sliding between two layers
leads to a shift of the moir\'{e} interference pattern in the real space.
The only exception is
when the two layers shares the same lattice vectors $\Vec{G}_i=\tilde{\Vec{G}}_i$,
where $\Vec{G}^{\rm M}_i$ vanishes so that
$\GVec{\tau}_0$ cannot be eliminated by shifting the origin.
In this case, the energy band actually becomes different
depending on the sliding vector $\GVec{\tau}_0$.
For the graphene bilayer case, in particular,
the expression of $U$ with $\Vec{G}_i=\tilde{\Vec{G}}_i$  becomes equivalent 
to that of regularly-stacked graphene bilayer with a interlayer sliding.
\cite{mucha2011strained,son2011electronic,koshino2013electronic}

\section{Conclusion}

We theoretically studied the interlayer interaction in general incommensurate bilayer systems 
with arbitrary crystal structures.
Using the generic tight-binding formulation,
we demonstrate that the interlayer coupling in the reciprocal space
is simply expressed in terms of a generalized Umklapp process.
We applied the formula to 
the incommensurate honeycomb lattice bilayer with a large rotation angle,
which cannot be treated as a long-range moir\'{e} superlattice,
and actually obtain the quasi band structure and density of states
within the first-order approximation.
Finally, we apply the formulation to the moir\'{e} superlattice 
where the two lattice structures are close, and derive the long-range effective theory
with a straightforward calculation.

\section*{ACKNOWLEDGMENTS}

This work was supported by JSPS Grant-in-Aid for Scientific Research No. 24740193
and No. 25107005.

\section*{References}

\bibliography{njp_moire}

\end{document}